\documentclass[12pt, manuscript]{aastex}
\usepackage{emulateapj5}

 \shortauthors{Strolger et al.}
 \shorttitle{SN~1999aw in Anonymous Faint Host Galaxy}

\begin{document}
{\let \uppercase \relax \title{THE TYPE Ia SUPERNOVA 1999aw: \\ 
A PROBABLE 1999aa--LIKE EVENT IN A LOW-LUMINOSITY HOST GALAXY}}    

 \author{L.-G.~Strolger,$^{1, 2}$ R.~C.~Smith,$^3$ N.~B.~Suntzeff,$^3$
 M.~M.~Phillips, $^4$ G.~Aldering,$^{5, 2}$ P.~Nugent,$^{5, 2}$
 R.~Knop,$^5$ S.~Perlmutter,$^5$ R.~A.~Schommer,$^3$ L.~C.~Ho,$^6$
 M.~Hamuy,$^6$ K.~Krisciunas,$^3$ L.~M.~Germany,$^{8, 2}$
 R.~Covarrubias,$^{9, 3}$ P.~Candia,$^3$ A.~Athey,$^1$
 G.~Blanc,$^{10}$ A.~Bonacic,$^{11}$ T.~Bowers,$^7$ A.~Conley,$^{5,
 12}$ T.~Dahl\'{e}n,$^{13}$ W.~Freedman,$^6$ G.~Galaz,$^4$
 E.~Gates,$^{14}$ G.~Goldhaber,$^{5, 12}$ A.~Goobar,$^{15}$
 D.~Groom,$^{5, 2}$ I.~M.~Hook,$^{16, 17}$ R.~Marzke,$^6$
 M.~Mateo,$^1$ P.~McCarthy,$^6$ J.~M\'{e}ndez,$^{18, 19}$
 C.~Muena,$^4$ S.~E.~Persson,$^6$ R.~Quimby,$^5$ M.~Roth,$^4$
 P.~Ruiz-Lapuente,$^{18}$ J.~Seguel,$^{20}$ A.~Szentgyorgyi,$^{21}$
 K.~von~Braun,$^1$ W.~M.~Wood-Vasey,$^{5, 12}$ and T.~York$^5$}

 \affil{$^1$University of Michigan, 830 Dennison Bldg., Ann Arbor, MI 
 48109-1090.\\loust@umich.edu}
 
 \affil{$^2$Visiting Astronomer, National Optical Astronomical
 Observatories, Cerro Tololo Inter-American Observatory.  \\{\scriptsize The National
 Optical Astronomical Observatories is operated by the Associated
 Universities for Research in Astronomy (AURA), Inc.  under cooperative
 agreement with the National Science Foundation.}} 
 
 \affil{$^3$National Optical Astronomical
 Observatories, Cerro Tololo Inter-American Observatory, Casilla 603, La
 Serena, Chile.} 
 
 \affil{$^4$The Observatories of the Carnegie Institution of
 Washington, Las Campanas Observatory, Casilla 601, La Serena, Chile.} 
 
 \affil{$^5$E. O. Lawrence Berkeley National Laboratory, One
 Cyclotron Rd., Berkeley, CA 94720.} 
 
 \affil{$^6$The Observatories of the Carnegie Institution of
 Washington, 813 Santa Barbara St., Pasadena, CA 91101.} 
 
 \affil{$^7$Steward Observatory, University of Arizona, Tucson, 
 AZ 85721.} 
 
 \affil{$^8$European Southern Observatory, Casilla 19001, 
 Santiago, Chile.} 
 
 \affil{$^9$Astronomy Department, University of Washington, Box 
 351580, Seattle, WA 98195-1580.} 
 
 \affil{$^{10}$CEA, DSM, DAPNIA, Centre d'Etudes de Saclay, 91191 
 Gif-sur-Yvette Cedex, France.} 
 
 \affil{$^{11}$Pontificia Universidad Cat\'{o}lica, Santiago, Chile} 
 
 \affil{$^{12}$Department of Physics, University of California, 
 Berkeley, CA 94720-7300.} 
 
 \affil{$^{13}$Stockholm Observatory, SCFAB, SE-106 91 Stockholm, Sweden.} 
 
 \affil{$^{14}$UCO/Lick Observatory, University of California, 
 Santa Cruz, CA 95064.} 
 
 \affil{$^{15}$Physics Department, Stockholm University, SCFAB,
 SE-106 91 Stockholm, Sweden.} 
 
 \affil{$^{16}$Institute for Astronomy, Royal Observatory, 
 Blackford Hill, Edinburgh EH9 3HJ, England, UK.} 
 
 \affil{$^{17}$Oxford University, Astrophysics, Keble Road, 
 Oxford OX1 3RH, England, UK.} 
 
 \affil{$^{18}$Department of Astronomy, University of Barcelona, 
 Marti\'{i} Franque's 1, Barcelona, E-08028, Spain.} 
 
 \affil{$^{19}$Isaac Newton Group, Apartado de Correos 321, Santa 
 Cruz de la Palma, Tenerife, Canary Islands, E-38780, Spain.} 
 
 \affil{$^{20}$Universidad de Concepci\'{o}n, Departmento de Fisica, 
 Casilla 4009, Concepci\'{o}, Chile.} 
 
 \affil{$^{21}$Harvard-Smithsonian Center for Astrophysics, 60
 Garden Street, Cambridge, MA 02138.}

\begin{abstract}
    SN~1999aw was discovered during the first campaign of the Nearby
    Galaxies Supernova Search ({\it NGSS}) project.  This luminous,
    slow-declining ($\Delta$$m_{15}(B)$$= 0.81\pm 0.03$) Type Ia
    supernova was noteworthy in at least two respects.  First, it
    occurred in an extremely low luminosity host galaxy that was not
    visible in the template images, nor in initial subsequent deep
    imaging.  Secondly, the photometric and spectral properties of
    this supernova indicate that it very likely was similar to the
    subclass of Type Ia supernovae whose prototype is SN~1999aa.  This
    paper presents the {\it BVRI} and {\it J$_{s}$HK$_{s}$}
    lightcurves of SN~1999aw (through $\sim100$ days past maximum
    light), as well as several epochs of optical spectra.  From these
    data we calculate the bolometric light curve, and give estimates
    of the luminosity at maximum light and the initial $^{56}$Ni mass. 
    In addition, we present deep {\it BVI} images obtained recently
    with the Baade 6.5-meter telescope at Las Campanas Observatory
    which reveal the remarkably low-luminosity host galaxy.
\end{abstract}

    
\section{Introduction}
 
 Type Ia supernovae (SNe~Ia) offer arguably the most precise method to
 measure cosmological distances.  Over the last ten years, these
 highly luminous explosions have been used to determine distances
 accurate to $7\%$, despite the fact that we know little about their
 progenitors.  These objects show considerable uniformity in their
 absolute {\it B} magnitudes at maximum light with an intrinsic
 dispersion of less than 0.4 mag (Hamuy et al.  1996b).  This scatter
 is greatly reduced by the application of empirical relations linking
 the luminosity at maximum to the width, or decay rate, of the
 lightcurve (Luminosity-Width Relations, or {\em LWR}s).  More
 luminous SNe~Ia decline in brightness after maximum at slower rates
 than less luminous SNe~Ia.  The $\Delta$$m_{15}(B)$ relation
 (Phillips 1993, Hamuy et al.  1996a, Phillips et al.  1999), which is
 a measure of the decay in the {\it B} band lightcurve from peak to 15
 days after peak, has reduced the scatter around the hubble law to
 less than 0.2 mag, making it a powerful tool in using SNe~Ia at high
 redshifts to investigate cosmological parameters.  Including
 reddening corrections further decreases the dispersion in the Hubble
 diagram to 0.14 mag (Phillips et al.  1999).  An equally effective
 method developed by Riess, Press, \& Kirshner (1996) uses the
 lightcurve shapes in multiple passbands to simultaneously estimate
 the SN~Ia luminosity and amount of extinction/reddening.  This MLCS
 method has demonstrated it can produce Hubble diagrams with
 dispersions of only 0.12 mag (6\% in distance).
 
 Attention is now being directed to the possible systematic errors
 involved in using these objects as high redshift distance indicators. 
 Part of the challenge lies in untangling the dispersion of SNe~Ia
 lightcurve widths from possible sub-populations of Type Ia
 supernovae.  With more discoveries of nearby supernovae in various
 host environments, and the development of new spectroscopic and
 photometric techniques for isolating these sub-populations, we may
 soon be able to determine $\Omega_{M}$ and $\Omega_{\Lambda}$ with
 lower systematic uncertainties as well as hone in on the progenitors
 of SNe~Ia.

 Over the past three years, the Nearby Galaxies Supernova Search Team
 {\it(NGSS)} has conducted successful search campaigns for supernovae
 of all types using the Kitt Peak National Observatory's 36-inch
 telescope and the Mosaic North camera (Mosaic I).  Each of our 5 - 8
 night campaigns have allowed us to search $\sim$ 250 fields (each
 nearly 1$\arcdeg$ square) along the celestial equator and out of the
 Galactic plane to limiting magnitudes of $R \sim 21$.  At the
 project's end, we had searched nearly 750 fields and discovered 42
 supernovae.  The goals of this project have been to understand
 supernova rates (for all SNe types) in both field and galaxy cluster
 environments, to investigate correlations of SN type with host galaxy
 environments, and to increase knowledge of observationally rare and
 peculiar SN types through increased detailed photometric and
 spectroscopic observations.  Further information concerning the {\it
 NGSS} project goals and methods will be described in a forthcoming
 paper (Strolger et al.  2003b, in preparation, hereafter referred to
 as Paper 2).

 SN~1999aw was one of the supernovae discovered during the first of
 the {\it NGSS} campaigns (Feb.  20 - Feb.  24, and Mar.  04 - Mar. 
 09, 1999) which was conducted in co-operation with the Supernova
 Cosmology Project ({\it SCP}; Aldering 2000, Nugent \& Aldering
 2000).  Initial discovery and confirmation images surprisingly showed
 a bright new object in a location which was devoid of galaxies on the
 template images taken only a few weeks earlier.  Spectroscopic and
 photometric evidence show SN~1999aw is a member of an intriguing
 subclass of Type Ia supernovae, characterized spectroscopically by
 SN~1999aa (Li et al.  2001).  These unusual supernovae have
 lightcurve shapes that are slightly different from those of normal
 Type Ia (Strolger et al.  2003a, also in preparation; see also
 Section~\ref{subsec:templates}).  These differences, although not yet
 completely understood, may be key in understanding not only the
 physical processes of this subclass, but of all SNe~Ia, and may place
 limits on models of Type Ia progenitors.
 
 In Section~\ref{sec:discovery} we discuss the discovery,
 confirmation, and classification of SN~1999aw.  In
 Section~\ref{sec:spectroscopy} we present several epochs of optical
 spectra and discuss their similarity to 1999aa-like supernovae.  In
 Section~\ref{sec:lcurves} we present the optical and infrared
 photometry, and the calibrations and corrections.  In
 Section~\ref{sec:analysis} we show the photometric similarity to
 1991T/1999aa SNe, determine the bolometric lightcurve, estimate the
 luminosity at maximum light, and estimate the initial $^{56}$Ni mass. 
 In Section~\ref{sec:galaxy} we present the photometry of the host
 galaxy and discuss the host environment.

\section{Discovery, Confirmation, and Initial Identification}
 \label{sec:discovery}
\begin{figure*}
    \epsscale{1.0} \plotone{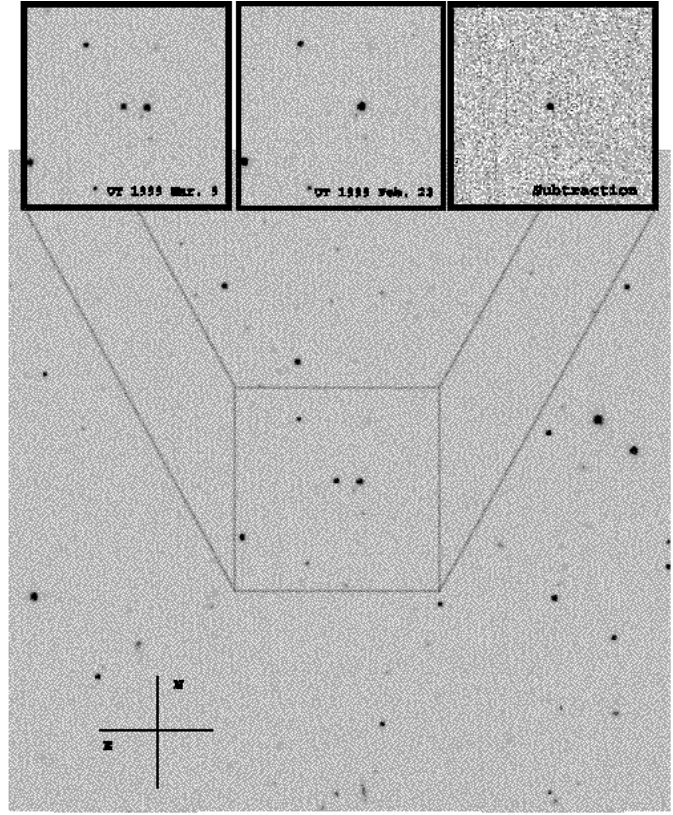} \caption{Discovery image of
    SN~1999aw taken on UT 1999 Mar.  9 on the KPNO 0.9-meter
    telescope.  Image is $\sim 6 \arcmin$ across ($0.423
    \arcsec$/pixel).  Superimposed at the top are the discovery image,
    the template image (taken UT 1999 Feb.  23), and the subtraction. 
    Algorithms used to produce the residual image are outlined briefly
    in Section~\ref{sec:discovery}.\label{fig:discovery}}
\end{figure*}

 SN~1999aw was discovered in search images obtained on UT 1999 Mar.  9
 with J2000.0 coordinates R.A. = 11h01m36.37s, Decl.  =
 $-$06\arcdeg06\arcmin31.6\arcsec\footnote{Based on the WCS, as
 determined from stellar registrations to the USNO2 Catalog.}.  No
 star was seen at its location in images obtained on UT 1999 Feb.  23
 (see Figure~\ref{fig:discovery} for the discovery image).  Our
 methods for the discovery of candidate supernovae are outlined in
 detail in Paper 2.  To summarize, for a given field, a pair of second
 epoch images were taken two weeks to two months after a template
 image was obtained.  These second epoch images were then aligned to
 the template image by stellar matching algorithms.  The images from
 the epoch with the better seeing were convolved to match that of the
 worst, and then scaled to be at the same flux level.  The template
 image was subtracted from the second epoch images to produce residual
 images that, in principle, contain only variable objects, transients,
 and cosmic rays, on a nearly zero-level background with noise.  The
 residual images were then automatically searched for candidates, with
 search parameter ranges set to eliminate cosmic rays and moving
 objects by requiring consistency between the pairs of second-epoch
 images.  As described in Paper 2, the routines used in the image
 subtractions for this first campaign were developed by the Supernova
 Cosmology Project for use in their high-z supernova searches
 (Perlmutter et al.  1997, 1999).

\begin{figure*}
    \epsscale{1.0} \plotone{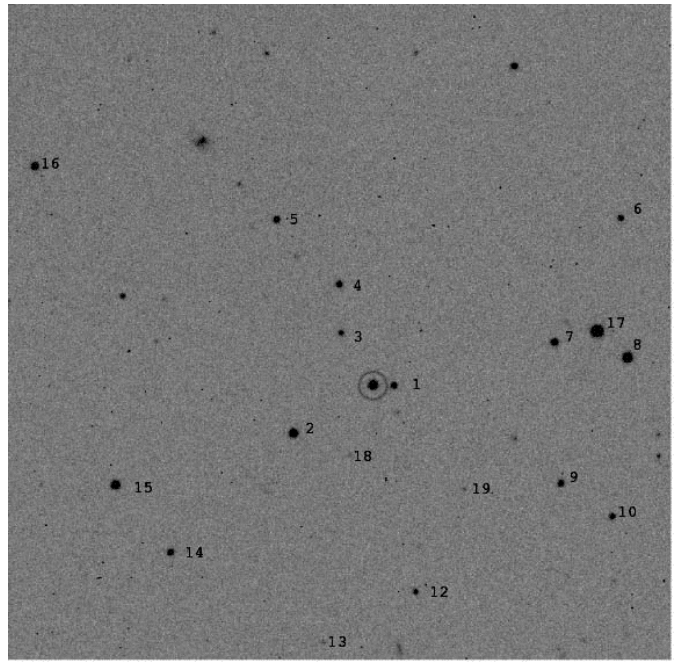} \caption{240 sec exposure of
    SN~1999aw (and local photometric sequence) in {\it B} from CTIO
    0.9-meter on UT 1999 Mar.  20.3.  Circle indicates SN. Image is
    $\sim 7 \arcmin$ across ($0.396 \arcsec$/pixel).  North is up,
    East is to the left.\label{fig:locals}}
\end{figure*}

 The candidate supernova was photometrically confirmed on UT 1999 Mar. 
 10 in direct images obtained by R. Covarrubias at the Cerro Tololo
 Inter-American Observatory (CTIO) 0.9-meter telescope.  The direct
 images showed that the object had not moved, within the limits of the
 seeing, in the time since its discovery.  Figure~\ref{fig:locals}
 shows SN~1999aw 11 days after discovery, and 3 days after maximum
 light.
 
\begin{figure*}
    \epsscale{1} \plotone{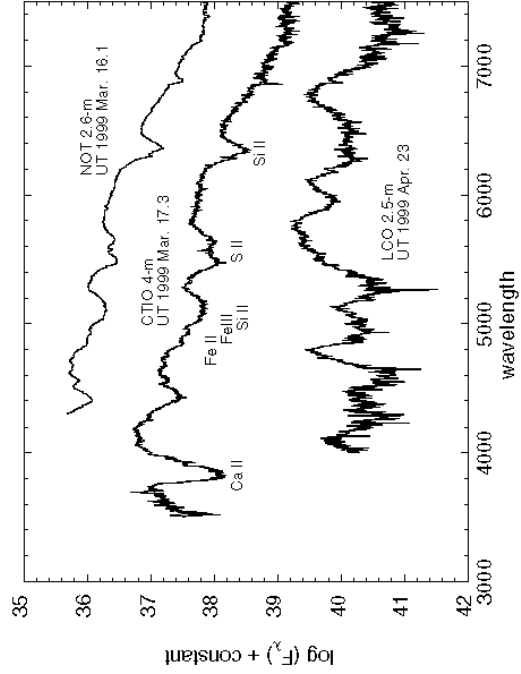} \caption{Spectral sequence of
    SN~1999aw.  Confirmation spectra from NOT (upper spectrum) and CTIO
    (middle spectrum) were obtained near maximum light.  Lower
    spectrum (From LCO) shows SN near initialization of the nebular
    phase.  This was used to determine the redshift to SN~1999aw of
    $z=0.038\pm0.001$.\label{fig:spectra}}
\end{figure*}

 The candidate was identified as a Type Ia supernova near maximum
 light from spectra obtained by A. Goobar, T. Dahlen, and I. Hook on
 UT 1999 Mar.  16.1 at the 2.6-meter Nordic Optical Telescope (NOT)
 using the Andalucia Faint Object Spectrograph and Camera
 (ALFOSC)\footnote{Observations were made with the Nordic Optical
 Telescope, operated on the island of La Palma jointly by Denmark,
 Finland, Iceland, Norway, and Sweden, in the Spanish Observatorio del
 Roque de los Muchachos of the Instituto de Astrofisica de Canarias.}. 
 Wavelength coverage was $\sim 4000-8000${\AA} with a resolution of
 700 (or 430 km/s).  A day later, L.-G. Strolger and R. C. Smith also
 obtained spectra at the CTIO 4.0-meter Blanco telescope (UT Mar. 
 17.3) using the R-C Spectrograph.  The effective wavelength coverage
 was 3400 - 7500 {\AA}, with an approximate resolution of 1000 (or 300
 km/s)\footnote{The 3K x 1K Loral CCD was used with the KPGL2 grating
 and the blue collimator.}.  From these spectra (presented in
 Figure~\ref{fig:spectra}), M. M. Phillips confirmed the initial
 identification by Goobar et al., and based on the very small ratio of
 Si~II$\lambda$5978 absorption relative to the Si~II$\lambda$6355 line
 (see Nugent et al.  1995a), suggested that SN~1999aw was likely to be
 a luminous, slow-declining SN~Ia.
 
\section{Optical Spectroscopy}
 \label{sec:spectroscopy}
 
 Although SNe~Ia generally show similar spectral evolution (e.g.,
 Branch et al.  1993), Nugent et al.  (1995a) showed that a spectral
 sequence exists for SNe~Ia which is analogous to the
 luminosity-decline rate relations.  Specifically, certain spectral
 features show a systematic variation as a function of
 $\Delta$$m_{15}(B)$.  Nugent et al.  argued that this spectral
 sequence is due primarily to the differences in effective temperature
 which are presumably correlated with the amount of $^{56}$Ni produced
 in the explosion.
 
 At maximum light, spectroscopically normal SNe~Ia exhibit strong
 Ca~II H and K features near 3750 {\AA}, and strong Si~II$\lambda$6355
 absorption which is blue-shifted by the high velocity of the
 expansion to appear near 6150{\AA} in the rest frame of the event
 (Minkowski 1940, Pskovskii 1969, Branch \& Patchett 1973).  Excellent
 examples of normal SNe Ia are SNe~1981B, 1989B, 1992A, and 1994D. In
 general, there are two groups of the spectroscopically ``peculiar''
 Type Ia SNe, also characterized near maximum light: SN~1991T-like
 events, and SN~1991bg-like events.  1991bg-like supernovae show wide
 absorption from 4150 - 4400 {\AA} due to Ti~II, and an enhanced 5800
 {\AA} feature (commonly attributed to Si~II) in relation to
 Si~II$\lambda$6355 (Filippenko et al.  1992, Leibundgut et al. 
 1993).  The 5800 {\AA} feature has been recently shown to be
 dominated by Ti~II absorption (rather than Si~II) in SN~1991bg-like
 SNe, and may even be considerably significant in spectroscopically
 normal SNe (Garnavich et al.  2001).  Alternatively, 1991T-like SNe
 have prominent Fe~II and Fe~III features at maximum light, but little
 or no Si~II, S~II, or Ca~II (Phillips et al.  1992).  A week after
 maximum light, however, the Si~II, S~II and Ca~II features develop,
 and the spectra become virtually indistinguishable from normal Type
 Ia.
 
 Nugent et al.  (1995a) showed that 1991bg-like events correspond to
 the low-temperature end of the SNe~Ia spectroscopic sequence (a fact
 supported by Garnavich et al.  2001), while 1991T-like events are
 associated with the highest effective temperatures.  Normal SNe~Ia
 inhabit the middle range of the sequence, where Ti~II absorption is
 weak, the ratio of Si~II$\lambda$6355 to Si~II$\lambda$5800 is
 relatively high, and the Ca~II H and K trough is strong.  Hence it
 may be more precise to refer to 1991bg-like and 1991T-like events as
 the extremes in a spectroscopic sequence rather than as ``peculiar''
 events.
 
 Li et al.  (2001) also document a variation of SN~1991T-like events
 for supernova spectra that resemble SNe 1999aa, 1998es, and 1999ac. 
 These SN~1999aa-like supernovae are similar to 1991T-like SNe
 supernovae, but show {\em some} Si~II$\lambda$6355 prior to maximum
 light (stronger than in 1991T-like SNe, but weaker than in normal
 Type Ia SNe), and clearly present Fe~II and Fe~III lines near maximum
 light.  They also exhibit strong Ca~II H and K lines.

\begin{figure*}
    \epsscale{1}
    \plotone{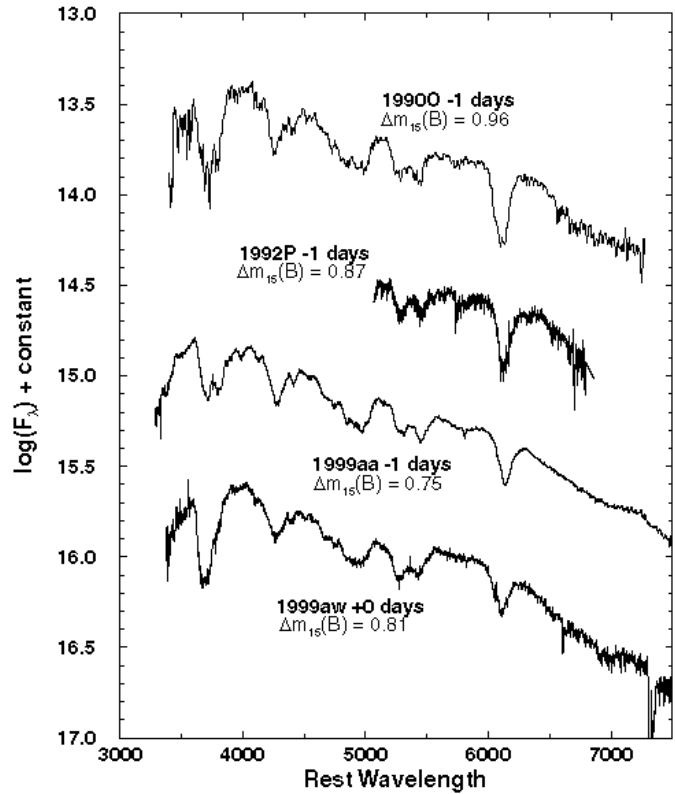}
    \caption{Comparison of SN~1999aw spectrum to other supernovae 
    near maximum light. SN~1999aw is more similar to SN~1999aa, than 
    to SNe 1990O or 1992P.\label{fig:comp}}
\end{figure*}

 The spectra of SN~1999aw obtained by Goobar et al.  (Mar.  16.1) and
 Strolger and Smith (Mar.  17.3) are displayed in
 Figure~\ref{fig:spectra}, and clearly show a SN~Ia at or near maximum
 light.  In Figure~\ref{fig:comp}, the Mar.  17.3 spectrum is compared
 with maximum-light spectra of four other SNe~Ia with similarly-slow
 decline rates.  SN~1990O was observed by Hamuy et al.  (1996c), and
 in spectra as early as 8 days before maximum clearly showed strong
 Si~II$\lambda$6355 absorption (Phillips et al., in preparation). 
 SN~1992P was also observed by Hamuy et al.  (1996c).  Although
 nothing is known about its spectrum a week or more before maximum,
 the spectrum plotted in Figure~\ref{fig:comp} also shows strong
 Si~II$\lambda$6355 absorption at maximum.  SN~1999aa is the prototype
 of the 1999aa-like events described in the previous paragraph. 
 Overall, the spectrum of this supernova (reproduced from Fig.  5 of
 Li et al.  2001) is very similar to those of SNe~1990O and 1992P,
 except that the Si~II$\lambda$6355 absorption is noticeably weaker. 
 At the bottom of Figure~\ref{fig:comp} we plot the spectrum of
 SN~1999aw obtained on Mar.  17.3.  The similarity to the spectrum of
 SN~1999aa is striking, with the only significant differences being
 the somewhat broader lines and stronger Ca~II absorption of
 SN~1999aw.  In section~\ref{sec:lcurves} we shall give further
 evidence for such a link based on the optical and IR light curves.
 
 On UT 1999 Apr.  23, M. M. Phillips obtained a spectrum of SN~1999aw
 using the duPont 2.5-m telescope at Las Campanas Observatory (see
 Figure~\ref{fig:spectra}).  The WFCCD was used in spectroscopic mode
 at a resolution of 630 (or 480 km/s) and the data covered the
 approximate wavelength range of 3750$-$9250{\AA}.  The date of
 observation corresponds to an epoch of 37 days after maximum light. 
 By this late epoch, SNe~Ia have begun their nebular phase, where
 their photospheres have essentially disappeared and their spectra
 become montages of blended emission features (Wells et al.  1994). 
 The peaks of these features, fortunately, do not change substantially
 with time, hence they can be used to determine the redshift of
 SN~1999aw relative to SNe~Ia with well-determined host galaxy
 redshifts and late epoch spectra.
 
 We compared the UT 1999 Apr.  23 spectrum to spectra of SNe 1992bc,
 1989B, and 1992A taken at epochs of +37 days, +36 days, and +36 days
 past maximum light respectively, and determined the relative shift
 between them.  Using the host galaxy redshifts given in
 NED\footnote{This research has made use of the NASA/IPAC
 Extragalactic Database (NED) which is operated by the Jet Propulsion
 Laboratory, California Institute of Technology, under contract with
 the National Aeronautics and Space Administration.}, we calculated
 the redshift for SN~1999aw of 0.0392, 0.0372, and 0.0380,
 respectively.  The mean result is $z=0.038\pm0.001$ which we will use
 throughout this paper.

\begin{figure*}
    \epsscale{1} \plotone{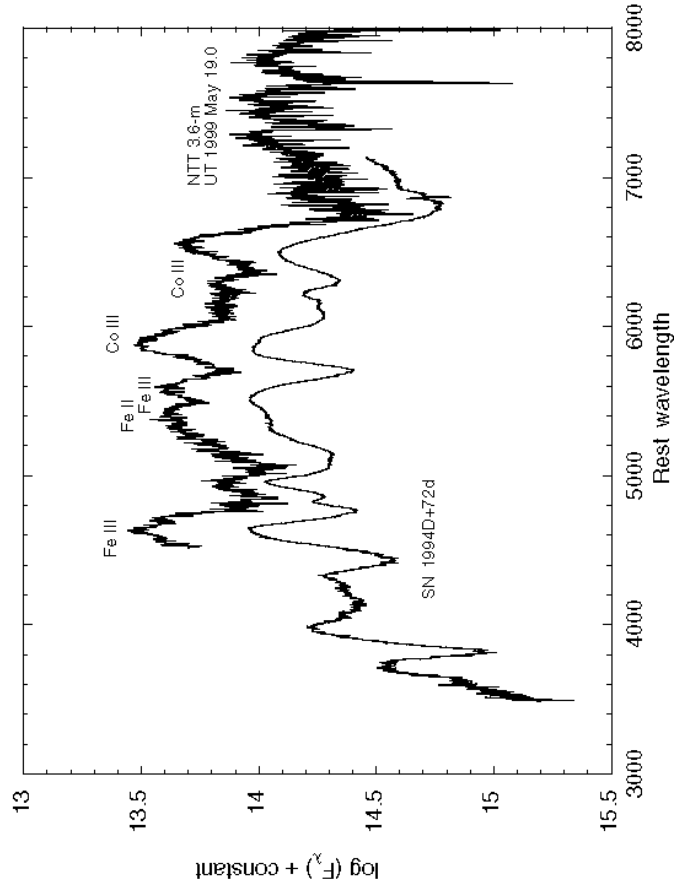} \caption{Additional late-time
    spectrum of SN~1999aw from NTT 3.6-m taken UT 1999 May 19.0. 
    Spectrum is shown at rest wavelength.  Nebular emission lines such
    as Fe~III$\lambda$4658 and Co~III$\lambda$5890 were used to
    calculate $z=0.037\pm0.002$, which is consistent with
    $z=0.038\pm0.001$ as derived from LCO 2.5-m spectrum (see
    Figure~\ref{fig:spectra}).  A spectrum of SN~1994D at $+$72 days
    from maximum is also shown for comparison.\label{fig:spectra2}}
\end{figure*}

 An additional late-time spectrum of SN~1999aw was obtained by M.
 Hamuy on UT 1999 May 19 using the NTT 3.6-m telescope at European
 Southern Observatory (see Figure~\ref{fig:spectra2}).  The EMMI was
 used in spectroscopic mode at a resolution of 600 (or 500 km/s) and
 an approximate wavelength range of 4600$-$8600{\AA}.  The date of
 this observation corresponds to an epoch of 64 days after maximum
 light, at which time the supernova was more ``nebular'' than in
 previous epochs, with more pronounced emission features. 
 Recalculating the redshift to SN~1999aw based on the centers of
 Fe~III$\lambda$4658 and Co~III$\lambda$5890 results in
 $z=0.037\pm0.002$, which is consistent with the redshift determined
 from the UT 1999 Apr.  23 spectrum.

\section{Photometry and lightcurves}\label{sec:lcurves}
\subsection{Optical Photometry}\label{subsec:optical}

 Our search techniques provided bulk SNe detections, and therefore we
 could schedule sufficient follow-up observations to produce
 well-sampled light curves.  Following the search campaign, we
 obtained several optical images using scheduled time on the CTIO
 0.9-meter, 1.5-meter, and 1.0-meter (YALO) telescopes.  The
 scheduling was planned such that the sampling would produce at least
 one data point every three nights, in each of the {\it BVRI} filters,
 for at least 30 days after the discovery.  We also obtained scheduled
 observations from various other telescopes.  Data were reduced (bias
 subtracted and flat-field corrected) using standard
 IRAF\footnote{IRAF is distributed by the National Optical Astronomy
 Observatories, which are operated by the Association of Universities
 for Research in Astronomy, Inc., under cooperative agreement with the
 National Science Foundation.} packages for reducing multiple and
 single amplifier data.  A shutter correction was measured for CTIO
 0.9-meter, and 1.5-meter data, and applied to short exposures ($< 20$
 seconds).
  
 The brightness of SN~1999aw, and its lack of host galaxy-light
 contamination, allowed accurate aperture photometry to be performed,
 without the need for late-time galaxy-light subtraction.  Using the
 DAOPHOT II package (Stetson 1992), data from photometric nights were
 compared with stars from tabulated standard fields (Landolt 1992). 
 Interactive iterative solutions were made to construct a growth curve
 and mean aperture correction for stars on each frame.  Solutions were
 then calculated for the coefficients of the linear color and airmass
 terms to convert from natural to standard apparent magnitudes (see
 Appendix, Tables~\ref{table:eqn} and \ref{table:terms}).  A set of
 local field standards stars was then produced around SN~1999aw (see
 Figure~\ref{fig:locals}), tied to the Landolt (1992) standard stars
 observed.  Table~\ref{table:locals} contains the sequence of local
 photometric standards, coordinate offsets in arcseconds from
 SN~1999aw (based on the WCS), and the photometric indices of these
 standards along with the mean errors.  Data of the SN from all other
 nights were compared to the data from the photometric nights using
 these standardized local field stars.  Table~\ref{table:observations}
 contains the epochs of observation of SN~1999aw, and the {\it BVRI}
 photometry along with the photometric errors.  These optical
 lightcurves are plotted in Figure~\ref{fig:lightcurve}.


 As expected from the identifying spectra, SN~1999aw did indeed dim
 after peak at a slower-than-average rate.  We performed a Monte Carlo
 simulation to randomly vary each data point within the photometric
 error 300 times, each time least-squares fitting a 3rd order
 polynomial to the peak of the {\it B} band lightcurve.  Averaging the
 solutions to the simulations led to the {\it B}-band peak of
 $B_{max}=16.88 \pm 0.01$ on JD 2451254.7$\pm 0.3$ (UT Mar.  17.2),
 and a decline rate within the first 15 days after maximum light of
 $\Delta$$m_{15}(B) = 0.80 \pm 0.03$.  Similar Monte Carlo simulations
 were performed on the {\it V, R} and {\it I} lightcurves, the results
 of which are summarized in Table~\ref{table:peaks}.  Using $z=0.038$
 and assuming $H_{o}=65$ km/s/Mpc, we derive the absolute magnitudes
 of $M_{B}=-19.48\pm0.11$, $M_{V}=-19.52\pm0.11$,
 $M_{R}=-19.52\pm0.11$, and $M_{I}=-19.04\pm0.12$, correcting only for
 Galactic extinction in the direction of SN~1999aw using values of
 $A_{B}=0.14$, $A_{V}=0.11$, $A_{R}=0.09$, and $A_{I}=0.06$ (Schlegel,
 Finkbeiner, \& Davis 1998).

\subsection{Infrared Photometry}\label{subsec:irphot}

 Infrared imaging of SN~1999aw was obtained over a 50-day period,
 beginning at the epoch of {\it B} maximum, using the Swope 1-meter
 and duPont 2.5-meter telescopes at Las Campanas Observatory (LCO),
 and using the VLT at European Southern Observatory.  These data were
 taken in the {\it J$_{s}$, H,} and {\it K$_{s}$}
 bandpasses\footnote{Subscript `s' is used to denote the modified
 filters used by Persson et al.  (1998).}.
 
 The LCO infrared images were reduced with IRAF using standard
 techniques.  Briefly, the steps consisted of 1) application of a
 correction for the slightly non-linear response of the NICMOS
 detector, 2) subtraction of dark images of the same exposure time as
 the supernova images, 3) division by a twilight flat field, 4)
 subtraction from each individual exposure of a sky image created from
 the dithered images of the supernova, and 5) shifting and summing the
 individual images to create a final ``mosaic'' image of the supernova
 field.  DAOPHOT II was used to obtain PSF photometry, construct
 growth curves, and to set up local standards using observations on
 photometric nights of the Persson et al.  (1998) standard stars.  As
 the Persson et al.  standards were established using the same
 instrument/detector/filters employed for the Swope telescope
 observations of SN~1999aw, no color corrections were applied to the
 supernova magnitudes.  The duPont IRC observations are probably also
 very nearly on the same system, but this may not be the case for the
 duPont CIRSI and VLT data.  It has been noticed that although the
 VLT/ISAAC transmission curves are fairly similar to the LCO filters
 in {\it H} and {\it K$_{s}$}, the {\it J$_{s}$} filter is
 significantly narrower and therefore the {\it J$_{s}$}-band
 observations can be quite different from the LCO system.  We hope to
 eventually derive appropriate transformations for the latter data to
 the Persson et al.  photometric system; in the absence of such
 information, no color corrections for these data are made in the
 present paper.

\begin{figure*}
    \epsscale{0.8}
    \plotone{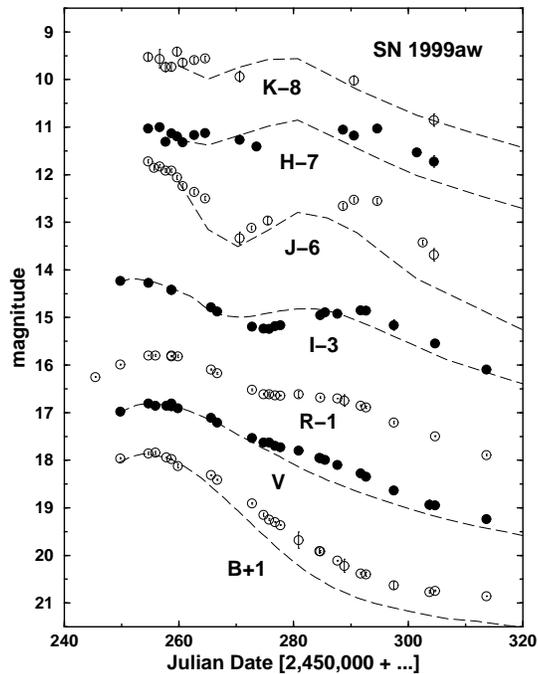}
    \caption{Optical and IR lightcurves of SN~1999aw with photometric 
    errors. Template curves for {\it B,V,} and {\it I} are from 
    the spectroscopically normal SN~1992al, which had a mid-range 
    decline rate of $\Delta$$m_{15}(B) = 1.11$ (Hamuy et al. 1996d). 
    Templates for {\it J$_{s}$, H,} and {\it K$_{s}$} from fiducial 
    curves of Elias et al. (1985b). Both the optical and IR templates 
    have been adjusted for a redshift of $z=0.038$.\label{fig:lightcurve}}
\end{figure*}

 Table~\ref{table:localir} lists the local standard star photometry
 (using the same numbering system as in the optical),
 Table~\ref{table:irobs} contains the epochs of observation of
 SN~1999aw with photometric errors, and Figure~\ref{fig:lightcurve}
 shows the resulting infrared light curves.

\subsection{K-Corrections and Time Dilation}\label{subsec:kcorr}
 
 At $z=0.038$, the redshift of SN~1999aw is large enough that it is
 important to correct the flux received in a given filter to that
 which would have been received if the event occurred at nearly zero
 redshift.  If spectroscopic data were available for each epoch a
 direct image was obtained, corrections could be made to the flux
 received in each passband, or K-corrections, by determining the
 change in brightness in the transmission curves of those passbands as
 the spectra are shifted from the observed reference frame to the rest
 frame.  Corrections would be applied as $m(z=0,\tau) = m(z,\tau) -
 K(z,\tau)$, where $\tau$ is the epoch from maximum light.  As this
 extensive spectral data is unavailable, we employ a method similar to
 that of Nugent et al.  (2002, in prep).  Here several epochs of
 spectra for SN~1999aw spanning the optical regime were used to
 produce the K-correction by manipulating the spectral energy
 distribution, producing synthetic photometry on a particular date to
 match the corresponding observed colors of SN~1999aw.  We interpolate
 and extrapolate these few K-terms to derive K-terms for all epochs of
 photometric observation.  The uncertainties from this method, based
 on the photometric data and the associated uncertainties for
 SN~1999aw, are less than 0.02 magnitudes for a given epoch.
 
 In at least {\it B} and {\it V}, the K-terms are in good agreement
 with the values produced in a pure application of the method of
 Nugent et al.  (2002), and with the values tabulated in Hamuy et al. 
 (1993) for a SN~Ia at $z=0.04$.  However, in {\it I} there is a
 significant difference from yet unpublished K-terms determined from
 other nearby SNe~Ia.  This difference is caused by the delay (or
 ``stretch'') in the {\it I}-band lightcurve of SN~1999aw, which is
 consistent with a possible trend among bright slow-declining
 1991T/1999aa-like SNe~Ia in which the {\it IJHK} lightcurves evolve
 at a slower rate than other SNe~Ia (see
 Section~\ref{subsec:colorcurves}).  This difference in photometric
 (and plausibly spectroscopic) evolution exemplifies the danger in
 applying K-corrections calculated from SNe~Ia with radically different
 decline rates.

%

 Applying the derived K-corrections to the photometric data for SN
 1999aw, and correcting the observed passage of time by a factor of
 $(1+z)^{-1}$ for time dilation, produced a rest frame apparent
 magnitude lightcurve that was not that different from the observed
 lightcurve, with $\Delta$$m_{15}(B) = 0.81\pm0.03$.  We have not
 attempted to make K-corrections for the near-IR data because we do
 not have IR spectra for SN~1999aw, nor are there yet sufficient
 libraries of IR spectra for other SNe~Ia.
 
\section{Analysis and Discussion}\label{sec:analysis}
\subsection{{\it B}-band Template Fits}\label{subsec:templates}
 
 The lightcurves of SN~1999aw are remarkable for how much slower they
 evolve compared to the template curves of a typical decline-rate SNe
 Ia (see Figure~\ref{fig:lightcurve}).  This is evident not only in
 the slow initial decline rate of $\Delta$$m_{15}(B) = 0.81$ (the mean
 decline rate of SNe Ia is $\Delta$$m_{15}(B)\sim 1.1$), but in the
 delay of the second maximum in {\it I, J$_{s}$, H,} and {\it
 K$_{s}$}.
 
 The {\it B}-band lightcurve is particularly interesting in that its
 ``shape'' is subtly different than that for typical SNe Ia.  One
 parameter LWR relations such as the $\Delta$$m_{15}(B)$ suggest that
 a {\it B}-band lightcurve template can be made to fit the {\it
 B}-band lightcurve of a SN by applying a time delaying ``stretch''
 factor to the epochs of observation, and the related magnitude offset
 (Perlmutter et al.  1997).  However, Strolger et al.  (2000 \& 2003a)
 show that {\it B}-band template lightcurves built from well-sampled
 spectroscopically normal SNe~Ia fit poorly to the {\it B}-band
 lightcurves of 1991T/1999aa-like supernovae, even when stretched to
 fit the observed lightcurve in early epochs (By ``normal'', we mean
 those those SNe~Ia which displayed strong Si~II$\lambda6355$
 absorption at least 5 days before {\it B}-band maximum).
 
 The Strolger et al.  (2000 \& 2003a) analysis was conducted on a few
 SNe that were 1) spectroscopically identified at least 5 days before
 maximum light, and 2) frequently observed with well sampled {\it
 B}-band lightcurves from just prior to maximum light to around the
 $+$80 day epoch.  The low-order mean difference between the template
 curve and the supernovae lightcurves from the beginning of the
 exponential phase (after the $+$25 day epoch) was determined for each
 supernova in the sample:

 \begin{equation}
     \delta_{ave} = \sum_{+25}^{+80} \frac{m(t) - m_{T}(t^{\prime})}{N}
 \end{equation}

 Observations made at some epoch, $m(t)$, were compared to the time
 stretched template curve, $m_{T}(t^{\prime})$, and then averaged over
 the number of observations ($N$) from the $+$25 day to $+$80 day
 epochs.
 
 Results of the analysis show that spectroscopically normal SNe Ia
 exhibit little to no difference from the template curve during the
 exponential phase, whereas 1991T/1999aa-like SNe Ia show a
 substantial overbrightness during this phase.  The analysis, when
 performed on SN~1999aw, produced a result consistent with that of the
 1991T/1999aa-like SNe.
 
\begin{figure*}
    \plotone{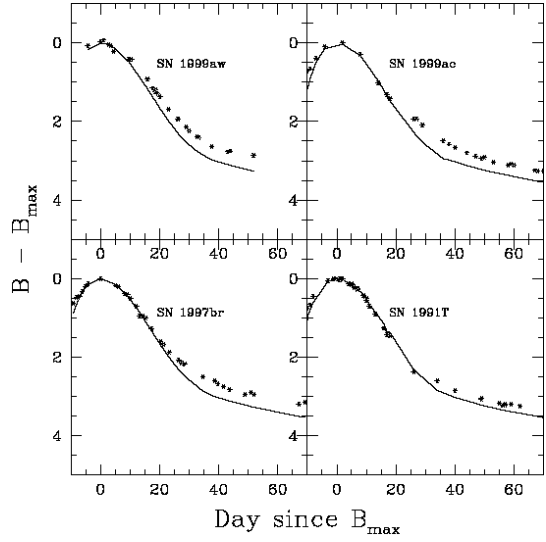}
    \caption{{\it B}-band lightcurves of four 1991T/1999aa-like 
    supernovae, scaled to {\it B}$_{max}$. Solid line is a Template 
    for spectroscopically normal Type Ia SNe, stretched along time 
    axis to fit observations between peak and $+$15 days past maximum 
    light.\label{fig:templates} }
\end{figure*}

 Figure~\ref{fig:templates} shows four SNe spectroscopically similar
 to SNe~1991T and/or 1999aa (including SN~1999aw), along with the
 normal SNe~Ia template, stretched in time to fit the lightcurves
 within the first 15 days past maximum light.  The template does not
 fit well to the data past around the $+25$ day epoch, and the data
 are systematically brighter than the template from that epoch.

 In the future, with more examples of 1991T/1999aa-like SNe, this
 analysis may lead to photometric method that, in addition to
 spectroscopy, indicate the possible 1991T/1999aa-like peculiarity of
 supernovae.

 \subsection{Color Curves}\label{subsec:colorcurves}

 SNe~Ia show a impressive uniformity in their intrinsic colors in late
 epochs after maximum light.  As Lira (1995) and Riess et al.  (1996)
 independently showed, SNe~Ia with
 $0.85\lesssim$$\Delta$$m_{15}(B)$$\lesssim 1.90$ and little or no
 reddening from their host galaxies have very uniform $B-V$ color
 evolution from $+$30 to $+$90 days after maximum light.  Krisciunas
 et al.  (2000) also observe uniformity in the {\it V}$-${\it Near IR}
 color evolution of spectroscopically normal SNe Ia with mid-range
 decline rates from $-$9 days to $+$27 days past maximum light.
  
\begin{figure*}
    \plotone{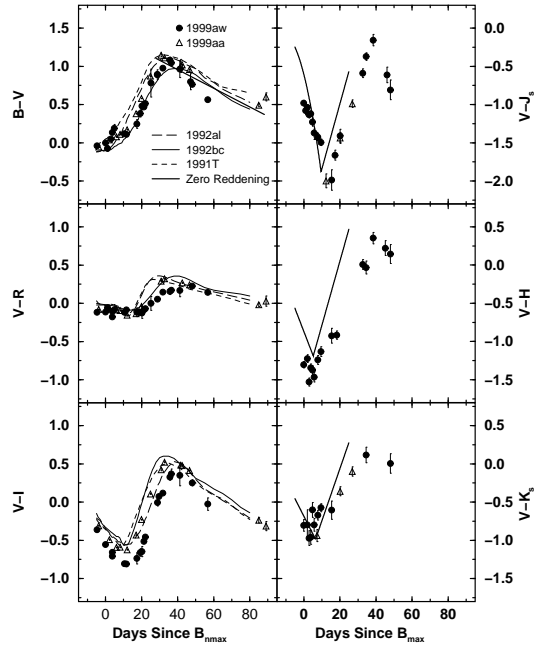} 
    \caption{Color curves of SN~1999aw with
    photometric errors. B$-$V curve includes Lira zero-reddening 
    relation (Phillips et al. 1999). V$-${\it Near IR} curves include 
    zero-reddening relations from Krisciunas et al. (2000) for Type ia 
    SNe with mid-range decline rates.\label{fig:colors}}
\end{figure*}

 Figure~\ref{fig:colors} shows the evolution in multiple colors of
 SN~1999aw, corrected for Galactic reddening assuming an excess of
 $E(B-V) = 0.032$ (Schlegel, Finkerbeiner, \& Davis 1998).  The
 left-hand side panels of Figure~\ref{fig:colors} show the optical
 color evolution, along with some example SNe (zero-reddening
 corrected) for comparison.  The solid line in the plot of the
 evolution of $B-V$ in this figure represents the zero-reddening
 least-squares fit derived by Lira (1995).  SN~1992al is included to
 show the evolution of a typical SNe Ia, while SN~1992bc is a
 slow-declining SNe Ia ($\Delta$$m_{15}(B) = 0.87$) but with normal
 pre-maximum light spectra.  SNe~1991T and 1999aa, the prototypical
 1991T/1999aa-like SNe, have color evolutions nearly parallel to the
 Lira (1995) line, thus showing that the uniformity holds for {\em
 some} spectroscopically extreme SNe~Ia.

 This color uniformity has proven useful as an indicator of host
 galaxy reddening, which has been important to revising the LWR
 relations (Phillips et al.  1999).  We have used the recipes in
 Phillips et al.  (1999) to determine the host galaxy reddening for
 SN~1999aw.  The analysis gave $E(B-V)_{tail}=-0.20\pm0.05$,
 $E(B-V)_{max}=0.12\pm0.10$, and $E(V-I)_{max}=-0.11\pm0.09$, which
 when averaged, resulted in a negative color excess, implying zero
 host galaxy reddening.  Clearly the negative value of $E(B-V)_{tail}$
 is heavily influenced by the anomalous $B-V$ color evolution of
 SN~1999aw; specifically, for most of the period from $+$40 days to
 $+$60 days SN~1999aw was considerably bluer than the Lira (1995)
 zero-reddening fit, and perhaps evolving with a steeper slope.
 
 As SN~1999aw was spectroscopically similar to SN~1999aa, we might
 have expected that its color evolution would be similar to SN~1999aa
 and/or SN~1991T, both of which appeared generally bluer than
 SN~1992bc in $V-R$ and $V-I$.  In SN~1999aw, the bluer $V-R$ and
 $V-I$ evolution were caused by the delay in the appearance of the
 second lightcurve maximum, which was more clearly seen in redder
 passbands (see Figure~\ref{fig:lightcurve}), and therefore is
 accentuated in the $V-R$ and $V-I$ curves.  SN~1999aw also appeared
 bluer than SN~1999aa and SN~1991T in $V-R$ and $V-I$, which perhaps
 was an effect of SN~1999aw having a much slower decline rate than
 both SN~1999aa and SN~1991T.
 
 The $V-J_{s}$ and $V-H$ curves for SN~1999aw are also considerably
 bluer than the zero-reddening fits from Krisciunas et al.  (2000) in
 the period after $+$10 days (right-hand side panels of
 Figure~\ref{fig:colors}).  The trend in the period prior to $+$10
 days in the $V-J_{s}$ curve also seems to be different than the fit
 from Krisciunas et al.  It appears that one could apply a ``stretch''
 to the Krisciunas lines to force them to fit the SN~1999aw data, that
 is to say there is an apparent delay in the $V-J_{s}$ evolution. 
 This again is due to the apparent delay of the second maximum in the
 {\it I, J$_{s}$, H,} and {\it K$_{s}$} lightcurves. Note, however, 
 that the $V-J_{s}$, $V-H$, and $V-K_{s}$ colors in this figure may 
 change somewhat once K-corrections for the IR bandpasses become 
 available.
 
\subsection{Bolometric lightcurve, Maximum Luminosity, and $^{56}$Ni 
Mass Estimate.}\label{subsec:bol}

 As nearly all of the bolometric luminosity of a typical Type Ia
 supernova is emitted in the range of 3000 to 10000{\AA} (Suntzeff
 1996), the integrated flux in the {\it UBVRIJ$_{s}$HK$_{s}$}
 bandpasses provides a reliable and meaningful estimate of the
 bolometric luminosity, which is directly dependent on the amount of
 nickel produced in the explosion.
 
 The {\it BVRIJ$_{s}$HK$_{s}$} data were used to calculate ``{\it
 uvoir}'' bolometric fluxes using the techniques described in Suntzeff
 (1996) and Suntzeff and Bouchet (1990).  A table of {\it
 UBVRIJ$_{s}$HK$_{s}$} data was made by linearly interpolating the
 {\it J$_{s}$HK$_{s}$} data to the optical dates.  For some of the
 missing optical and {\it J$_{s}$} and {\it H} data, we added
 photometry based on spline fitting of the data to the date of the
 missing data.  We have added on {\it U} data because the optical
 ultraviolet adds significant flux to the early time bolometric light
 curve.  There is little high-quality {\it U} data for Type Ia
 supernovae due to the generally poor ultraviolet sensitivity of the
 present generation of CCDs.  We have instead relied on $U-B$
 photometry from photoelectric measurements.  For dates past $+$9 days
 from {\it B} maximum, we have used the $U-B$ data of SN~1972E from
 Lee et al.  (1972) and Ardeberg \& de Groot (1973).  For dates before
 $+$9 days from {\it B} maximum we have used the $U-B$ data for
 SN~1980N from Hamuy et al.  (1991), and for SN~1981B compiled by
 Cadonau \& Leibundgut (1990).  The $U-B$ for these supernovae were
 corrected to the reddening of SN~1999aw using the reddening values in
 Phillips et al.  (1999) and a value of $E(U-B)/E(B-V)=0.72$ from
 Cardelli et al.  (1989).  The {\it U} photometry of SN~1999aw was
 then estimated from spline fits to the $U-B$ data of these supernovae
 combined with our {\it B} data of SN~1999aw.

 We then converted the broadband magnitudes to equivalent
 monochromatic fluxes at the effective wavelengths of Vega (Bessell
 1979, 1990; Bessell \& Brett 1988).  A magnitude scale of
 $(U,B,V,R,I)=0.03$ and $(J,H,K)=0.0$ was used for Vega.  The
 monochromatic fluxes were then scaled to the magnitude of the
 supernova, dereddened by $E(B-V)=0.032$ using the reddening law of
 Cohen et al.  (1981), and corrected to intrinsic fluxes using a
 distance modulus of 36.28 based on a Hubble flow with a Hubble
 constant of 63.3 km/s/Mpc (Phillips et al.  1999).  These fluxes were
 then integrated using a simple trapezoidal integration.  We added on
 a Rayleigh-Jeans extrapolation to zero frequency to the reddest flux
 point.  We extrapolated to the ultraviolet by adding a flux point at
 3000{\AA} with zero flux.

\begin{figure*}
    \epsscale{1.0} \plotone{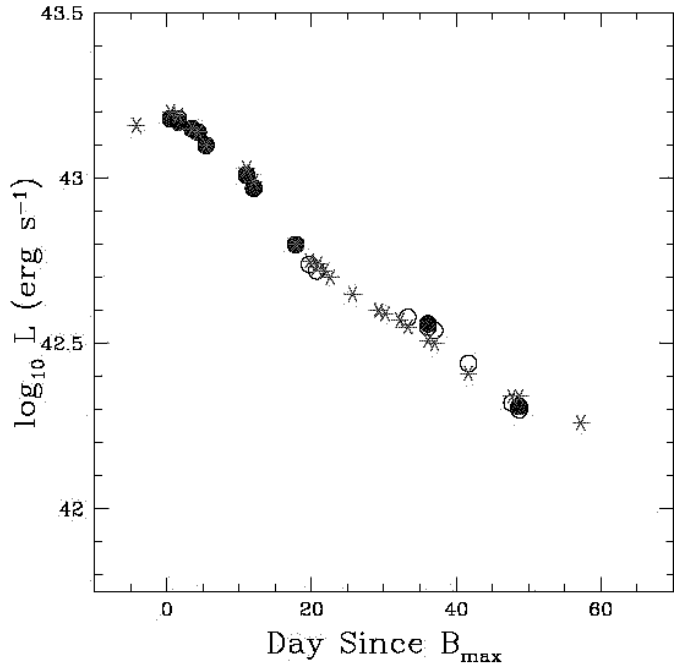} \caption{The bolometric
    lightcurve of SN~1999aw, constructed from the integrated flux in
    the {\it B, V, R, I, J$_{s}$, H,} and {\it K$_{s}$} passbands, and
    corrected for time dilation.  Stars show the bolometric curve in
    UBVRI, while open circles are in UBVRIJ$_{s}$H, and filled circles
    are in UBVRIJ$_{s}$HK$_{s}$.\label{fig:bolcurve}}
\end{figure*}

 We correct the derived bolometric lightcurves for time dilation
 effects, and plot them in Figure~\ref{fig:bolcurve}.  The {\it UBVRI}
 and the {\it UBVRIJ$_{s}$HK$_{s}$} integrations track each other
 well, except that the inflection points around days 20-45 are more
 pronounced in the {\it UBVRIJ$_{s}$HK$_{s}$} integrations.  Similar
 inflection points were noted by Suntzeff (1996) and are indicative of
 significant flux redistributions which may be related to rapid
 changes in the wavelength dependence of the opacities (Pinto \&
 Eastman 2000 a \& b).  The peak bolometric luminosity is about
 $L_{bol} = 1.51 \times 10^{43}$ erg s$^{-1}$.

 It is fairly straightforward to derive the nickel mass produced in
 the explosion from the bolometric luminosity at peak.  At maximum
 light, photons escape the surface at a rate which is equal to the
 radioactive energy input produced primarily by the $^{56}$Ni decay,
 and thus it is also related to the $^{56}$Ni synthesized in the
 explosion (Arnett 1982, Nugent et al. 1995b, Pinto \& Eastman 2000a).  

 Contardo et al.  (2000) calculate the luminosity and nickel mass for
 several SNe~Ia from {\it UBVRI} bolometric peak fluxes.  Using the
 same method, and assuming a rise time of 17 days to bolometric peak
 for consistency with Contardo, et al., we derive an initial nickel
 mass of $M_{Ni} = 0.76 M_{\odot}$ for SN~1999aw.  This is brighter
 and more nickel massive than many of the normal Type Ia SNe discussed
 in Contardo, et al., and it is comparable in brightness and nickel
 mass to SN~1991T (see Table~\ref{table:masses}).  However as they
 note, a number of the SNe~Ia in their study do not have {\it U}-band
 data available at peak, and therefore they have developed a
 correction curve based on data from the very well-sampled SN~1994D.
 Although this SN was spectroscopically normal, it did have some
 rather unusual features, including an unusually blue $U-B$ color at
 maximum light.  Additionally, Riess et al.  (1999) have shown that
 the characteristic rise time of SNe~Ia is $19.5\pm0.2$ days, and that
 brighter and slower declining SNe~Ia have longer rise times.  For the
 peak magnitude and decline rate observed in SN~1999aw, we expected a
 rise time of $\sim20$ days.  Using this value, we derive an
 alternative initial nickel mass of $M_{Ni}=1.07 M_{\odot}$.
 
 There are a number of additional uncertainties in both the calculated
 luminosity, and the derived nickel mass.  The first is the assumption
 that more than 80\% of the true bolometric light is emitted in the
 optical regime, that less than 10\% can be expected in the UV below
 3200{\AA}, and that no more than 10\% (in early epochs) from {\it
 JHK} (Suntzeff 1996, Elias et al.  1985, Contardo et al.  2000).  In
 comparing our derived {\it UBVRIJ$_{s}$HK$_{s}$} and {\it UBVRI}
 bolometric fluxes, we find the IR contribution to be only a few
 percent ($\sim 1\%$ at early epochs, $\sim 5\%$ after 35 days past
 $B_{max}$).  We have not accounted for the space ultraviolet flux.
 
 Also, as we will discuss further in Section~\ref{sec:galaxy}, we do
 not have much information about the host galaxy of SN~1999aw. 
 Although we have compensated for extinction and reddening due to our
 own galaxy, it difficult to do the same for the host galaxy. 
 However, as the $B-V$ color of SN~1999aw is nearly zero at maximum
 light, as it would be for an unreddened Type Ia SN (Phillips et al. 
 1999; Garnavich et al.  2001, in Figure 15), we have assumed that the
 extinction due to the host must be negligibly small.

\section{The Host Galaxy of SN~1999aw}\label{sec:galaxy}

\begin{figure*}
     \epsscale{1.0}
     \plotone{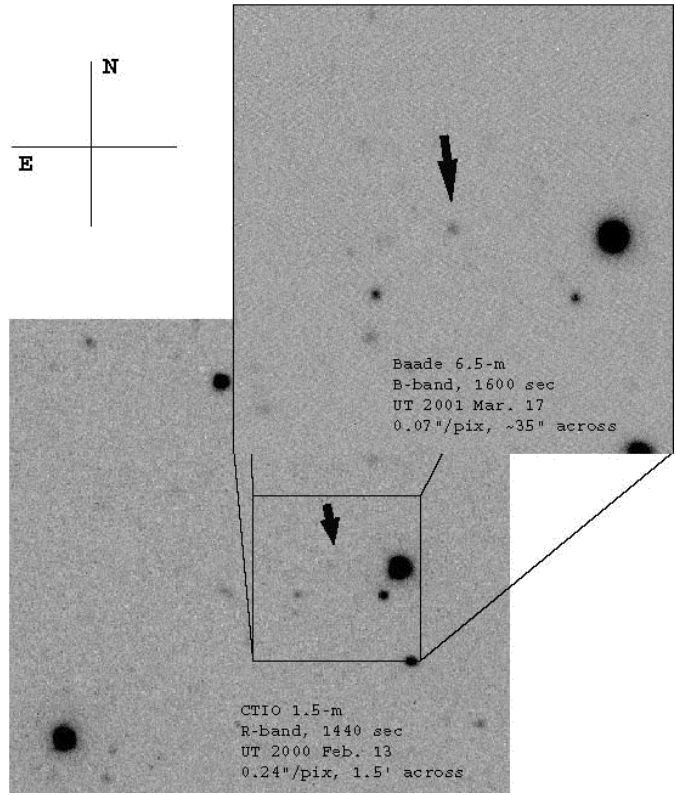}
     \caption{Deep imaging of the SN~1999aw field. Host is visible in 
     the Magellan 6.5-meter data. \label{fig:hostgxy}}
\end{figure*}

 Another approach to understanding SN~1999aw can be to investigate its
 host galaxy.  The metallicity of the progenitor may be related to the
 environment of the event (Hamuy et al.  1995).  However, the apparent
 low luminosity of the galaxy has made it very difficult to study.  As
 earlier stated, there was no obvious host galaxy in the template
 images nor in subsequent photometry.  On UT 2000 Feb.  12, L.
 Strolger, P. Candia, J. Seguel, and A. Bonacic\footnote{Seguel and
 Bonacic participated in the 2000 Research Experiences for
 Undergraduates (REU) and Pr\'{a}ctica de Investigat\'{i}on en
 Astronom\'{i}a (PIA) program at CTIO.} obtained deep images of the
 SN~1999aw field using the CTIO 1.5-meter telescope.  Pairs of long
 exposures were taken in the same standard {\em BVRI} filters as were
 used in the photometry of SN~1999aw (The combined exposure times were
 1200 seconds for B, V, and I, and 1440 for R).  Zero points in each
 filter were found by comparison of standards fields taken on the same
 evening, with tabulated standard magnitudes.  No clear detection (to
 a 3-$\sigma$ level) was made of a galaxy at or near the position of
 SN~1999aw in these deep images (see Figure~\ref{fig:hostgxy}).

 A second attempt to detect the host was made on UT 2001 Mar.  17 by
 A. Szentgyorgyi, and on UT 2001 Mar.  18 by M. Mateo, A. Athey, and
 K. von Braun, both using the Baade 6.5-meter telescope at Las
 Campanas Observatory and the Magellan Imaging camera (MagIC).  They
 obtained deep exposures for a combined 1600 sec in {\it B}, 1400 sec
 in {\it V}, and 900 sec in {\it I} of the SN~1999aw field.  These
 deep images did reveal a resolved galaxy, with FWHM of 0.6\arcsec,
 only slightly broader than the seeing (0.42\arcsec).  The central
 brightness (within 1.5{\arcsec} radii) were $B = 24.2 \pm 0.2$, $V =
 23.8 \pm 0.2$, and $I = 24.1 \pm 0.9$ magnitudes.  The radial growth
 curve for the galaxy was very steep in all passbands, and heavily
 contaminated by background noise in the image.  Therefore, we assume
 the 1.5{\arcsec} magnitudes are the best estimates of the total light
 in the galaxy.  Absolute values of $M(B)_{T}=-12.2\pm0.2$,
 $M(V)_{T}=-12.5\pm0.2$, and $M(I)_{T}=-12.2\pm0.9$ were calculated
 assuming $H_{o}=65$ km/s/Mpc and correcting only for Galactic
 extinction.

 Coincidentally, the low luminosity dwarf galaxy IC~4182 hosted type
 Ia supernova SN~1937C, also with slow decline rate
 ($\Delta$$m_{15}(B)=0.87$).  Although Branch et al.  (1993) indicate
 that spectra from Minkowski (1939 \& 1940) of SN~1937C seem to show a
 normal Type Ia SN, there may be some question as to whether or not
 the Si~II$\lambda$6355 feature was relatively weak at early epochs,
 possibly indicating a similarity to SN~1999aa.  Using the measured
 CCD photometry of IC~4182 from Makarova (1999), and the Cepheid
 distance modulus for the galaxy calculated by Saha, et al.  (1994)
 and correcting for Galactic extinction, we determined its absolute
 magnitude to be $M(B)_{T}=-16.4\pm0.02$, $M(V)_{T}=-16.8\pm0.2$, and
 $M(I)_{T}=-17.6\pm0.2$.  The SN~1999aw host galaxy is {\bf
 significantly fainter} (by 4.2, 4.3, and 5.5 magnitudes in {\it B},
 {\it V}, and {\it I}, respectively).

 The error in these observations, especially in {\it I}, make it
 difficult to estimate color differences for the host galaxy, however,
 using the magnitudes above yields $(B-V)_{o}=0.31\pm0.28$ and
 $(V-I)_{o}=-0.35\pm0.92$.  Assuming $(B-V)_{o}$ better exemplifies
 the color trend, the galaxy would appear to be fairly blue in
 comparison to dwarf galaxies in the local group (see Table 4 and Fig. 
 3 of Makarova 1999).
 
 Initial studies of SNe~Ia characteristics and their relation to their
 host galaxies have been conducted by Ivanov, Hamuy, \& Pinto (2000)
 and Hamuy et al.  (2000).  Their studies show:
 
 \begin{itemize}
     \item The distribution of SNe~Ia decline rates (and therefore
     brightness) changes considerably with host galaxy morphological
     type.  Earlier Hubble type galaxies (such as ellipticals and S0
     spirals) produce faint fast-declining SNe~Ia, and late type
     galaxies (Sa$-$Irr) tend to produce bright slow-declining SNe~Ia.
     
     \item SNe~Ia decline rates also change with $B-V$ color of the
     host galaxy.  Bright SNe~Ia occur more frequently in bluer
     environments.
     
     \item Bright SNe~Ia occur more frequently in less luminous
     galaxies.
 \end{itemize}

 It is difficult, however, to determine how these color$-$luminosity
 correlations relate to age and metallicity effects.  Integrated
 galaxy luminosities seem to correlate with metallicity (Henry \&
 Worthy 1999), and both the sample of galaxies from Hamuy et al. 
 (2000) and the sample of dwarf galaxies from Makarova (1999) seem to
 indicate that the least luminous galaxies are also the bluest.  As
 bluer galaxies are also younger, perhaps the brightness distribution
 of SNe~Ia with $B-V$ color is an age effect, suggesting that younger
 environments produce bright slow-declining SNe~Ia.  Alternatively, it
 is plausible that the SNe~Ia brightness$-$color correlation is just a
 reflection of the SNe~Ia brightness$-$host galaxy luminosity trend,
 which may be a metallicity effect indicating that metal-poorer
 environments produce the brightest SNe~Ia.

 Nonetheless, observations of SN~1999aw and its host galaxy are
 consistent with these trends in that the host was faint and blue
 [among the faintest 10\% in $M(B)$, and the bluest 15\% in $(B-V)$ of
 the Makarova (1999) sample], and that it produced a bright
 slow-declining SN~Ia.

\section{Summary}\label{sec:summary}

 Our photometric and spectroscopic study of SN~1999aw indicate that
 SN~1999aw was probably a 1999aa-like event.  The light curve decline
 rate is among the slowest observed at $\Delta$$m(B)_{15} = 0.81 \pm
 0.03$.  At the redshift of $z = 0.038$, it is also among the
 brightest Type Ia, with $M_{B}=-19.45\pm0.11$, $M_{V}=-19.50\pm0.11$,
 $M_{R}=-19.38\pm0.11$, and $M_{I}=-18.97\pm0.12$.  Although luminous,
 these magnitudes are slightly less (by a few tenths of a magnitude)
 than one might expect from the trends of absolute magnitude versus
 decline rate or $\Delta$$m_{15}$ (See Figure 11 of Krisciunas et al. 
 2001).  Perhaps this suggests that there is some host galaxy
 extinction that needs to be accounted for.  Or it is possible that
 SN~1999aw is simply slightly less luminous than expected, and
 therefore may be indicating a possible downward curve to the
 relations in Fig.  11 for the slowest-declining SNe Ia (similar to
 the downward curve recently determined by Garnavich et al.  (2001)
 for fast-declining SN 1991bg-like SNe~Ia).
 
 It is thought that the brightness and decline rate of Type Ia are
 dependent on the metallicity of the progenitor, and additionally on
 the opacity in the atmosphere of the event (Hoflich et al.  1998,
 Pinto and Eastman 2000a \& b, Mazzali et al.  2001).  The derived
 luminosity of $1.51 \times 10^{43}$ erg s$^{-1}$, and $^{56}$Ni mass
 of $0.76 M_{\odot}$ for SN~1999aw are both relatively high, but are
 consistent with what can be expected for a SN Ia with the observed
 decline rate, as inferred from the trend between $\Delta$$m_{15}(B)$
 and $^{56}$Ni mass evident in Table~\ref{table:masses}.
  
 The observations made at the Baade 6.5-meter have provided some very
 interesting information on the host galaxy of SN~1999aw.  The derived
 absolute magnitudes seem to indicate that it is among the
 intrinsically faintest and bluest dwarf galaxies observed.  This may
 indicate that the galaxy consists of either a young population of
 stars, or is a fairly metal-poor environment.  It will be necessary
 to eventually obtain deeper images with large telescopes such as the
 Magellan 6.5-meter or the VLT to obtain sufficient signal-to-noise to
 put more accurate constraints on not only the colors of this galaxy,
 but its structure as well.  The discovery of SN~1999aw exemplifies an
 advantage of magnitude limited field searches over galaxy-targeted
 searches.  It is presently unclear how many supernova similar to
 SN~1999aw occur at low redshift, and with the biases associated with
 current targeted surveys we are less likely to detect them.  As more
 magnitude limited field searches commence, relations between low
 luminosity (and probably low metallicity) galaxies and the supernova
 types that occur in them can be accurately determined.  This will not
 only place constraints on Type Ia supernova progenitor models, but
 help to clarify the possibility of supernova ``evolution'' which may
 explain the difference in SN~1991T-like event rates between the low
 and high-$z$ surveys (Li et al.  2001).

 \acknowledgements
 This work is dedicated to the memory of Robert A. Schommer, whose
 contributions we valued, and whose friendship we will miss.  LS is
 very grateful to Cerro Tololo Inter-American Observatory and its
 staff for providing a office, exceptional support, mentorship, and an
 unequivocal learning experience during his stay there from 1998 to
 2000.  Additional acknowledgements to B. Schaefer, C. Baiyln, and S.
 Tourtellotte for the use of YALO and YALO data, and to A. Block, M.
 Block, P. Challis, J. Krick, H. Mathis, and A. Soderberg for their
 assistance is searching through countless images in search of
 supernovae.
 
\appendix
 \section{Photometric Transformation Equations}
 Optical aperture photometry was performed using the DAOPHOT II
 package (Stetson 1992).  14{\arcsec} diameter apertures were used and
 were corrected by an iterative growth curve to a nearly infinite
 diameter.  Standard fields were observed on 8 photometric nights on
 the CTIO 0.9-meter, and 7 photometric nights on the CTIO 1.5-meter. 
 Caution was taken to observe fields at various airmasses in all
 filters.  The data from 50 stars were compared to tabulated values
 (Landolt 1992) and solutions were found for the coefficients of the
 linear color and airmass terms in the photometric transformation
 equations (see Table~\ref{table:eqn} and \ref{table:terms}).  The
 equations were then used to transform the natural magnitudes of the
 local standard stars to standard magnitudes (see
 Table~\ref{table:locals}).

\clearpage

\begin{deluxetable}{cccccccccccc}
    \tabletypesize{\scriptsize}
    \tablewidth{0pc}
    \tablenum{1}
    \tablecaption{Local Standard Indices\label{table:locals}}
    \tablehead{
    \colhead{Star} & \colhead{E (\arcsec)} & \colhead{N (\arcsec)} &
    \colhead{V} &\colhead{B$-$V}& \colhead{V$-$R} & \colhead{R$-$I} & 
    \colhead{V$-$I} & \colhead{N$_{B}$}& \colhead{N$_{V}$}& 
    \colhead{N$_{R}$}& \colhead{N$_{I}$}}
    \startdata
1&$-$13.5&0.6&17.138 (0.015)&1.346 (0.033)&0.861 (0.022)&0.846 (0.022)&1.707 (0.020)&23&20&21&21\\
2&49.7&$-$30.0&16.67 (0.014)&1.95 (0.023)&0.373 (0.020)&0.339 (0.023)&0.712 (0.022)&23&20&19&21\\
3&19.7&33.1&18.477 (0.028)&0.7 (0.049)&0.406 (0.038)&0.39 (0.038)&0.795 (0.039)&17&16&15&17\\
4&20.4&63.4&17.587 (0.021)&0.856 (0.035)&0.486 (0.027)&0.433 (0.024)&0.919 (0.027)&23&22&19&21\\
5&58.9&104.2&17.83 (0.019)&0.728 (0.035)&0.428 (0.028)&0.4 (0.029)&0.828 (0.028)&22&19&18&20\\
6&$-$154.8&103.7&18.178 (0.028)&0.691 (0.045)&0.394 (0.041)&0.378 (0.038)&0.772 (0.037)&19&18&17&17\\
7&$-$112.5&26.6&17.38 (0.018)&0.555 (0.028)&0.372 (0.026)&0.378 (0.025)&0.75 (0.024)&21&20&19&19\\
8&$-$157.6&16.1&15.65 (0.010)&0.961 (0.018)&0.547 (0.014)&0.476 (0.012)&1.024 (0.012)&17&16&15&17\\
9&$-$116.0&$-$62.1&17.488 (0.018)&0.938 (0.033)&0.526 (0.026)&0.457 (0.024)&0.983 (0.025)&20&19&18&18\\
10&$-$147.8&$-$83.4&18.201 (0.027)&0.59 (0.043)&0.358 (0.039)&0.351 (0.039)&0.708 (0.037)&21&20&19&19\\
11&$-$122.9&$-$196.0&15.632 (0.017)&1.019 (0.027)&0.411 (0.025)&0.377 (0.024)&0.788 (0.023)&12&11&10&10\\
12&$-$25.6&$-$129.3&18.529 (0.028)&0.773 (0.048)&0.463 (0.040)&0.442 (0.041)&0.905 (0.041)&19&18&17&17\\
13&51.6&$-$190.1&16.958 (0.016)&0.648 (0.025)&0.393 (0.024)&0.384 (0.025)&0.777 (0.023)&17&16&15&15\\
14&126.2&$-$104.2&18.092 (0.026)&0.45 (0.038)&0.326 (0.040)&0.321 (0.043)&0.647 (0.040)&19&16&17&17\\
15&160.3&$-$61.6&16.279 (0.012)&2.387 (0.021)&0.515 (0.021)&0.452 (0.024)&0.967 (0.021)&17&14&15&15\\
16&208.7&138.7&17.369 (0.015)&0.576 (0.023)&0.356 (0.026)&0.346 (0.029)&0.702 (0.025)&9&9&8&7\\
17&$-$139.3&33&15.052 (0.009)&0.559 (0.016)&0.334 (0.013)&0.333 (0.013)&0.667 (0.013)&21&20&19&19\\
18&14.0&$-$43.7&20.951 (0.061)&0.672 (0.093)&2.037 (0.073)&0.579 (0.062)&2.616 (0.077)&4&5&5&5\\
19&$-$56.1&$-$64.6&19.206 (0.043)&1.641 (0.067)&1.031 (0.050)&0.722 (0.040)&1.753 (0.052)&2&3&3&3\\
\enddata
    \tablecomments{Local photometric sequence with photometric
    error.  Offsets are in arcseconds from SN~1999aw (R.A. = 11h01m36.37s, Decl.  =
 $-$06\arcdeg06\arcmin31.6\arcsec) to star. Coordinates are 
    determined by comparison to the USNO2 Catalog.}
\end{deluxetable}

\begin{deluxetable}{ccccccc}
    \tabletypesize{\footnotesize}
    \tablenum{2}

    \tablewidth{0pc} \tablecaption{{\it BVRI} Aperture 
    Photometry\label{table:observations}} \tablehead{
    \colhead{JD} & \colhead{Telescope} & \colhead{Observer} &
    \colhead{B} & \colhead{V} & \colhead{R} & \colhead{I}}
    \startdata

    245.40& KPNO 0.9-m& NGSS Team&\nodata&\nodata&17.251
    (0.010)&\nodata\\
    249.76& YALO & Service &16.961 (0.016)& 16.980 (0.008)&
    16.990 (0.009)& 17.229 (0.027)\\ 254.66& YALO & Service
    &16.858 (0.023)& 16.806 (0.015)& 16.800 (0.007)& 17.272 (0.015)\\
    255.88& Lick 40-inch& Quimby &16.838 (0.023)& 16.856
    (0.028)& 16.797 (0.020)& \nodata \\ 257.80& CTIO 0.9-m &
    Strolger &16.941 (0.007)& 16.855 (0.007)& \nodata & \nodata \\
    258.66& YALO & Service &16.972 (0.005)& 16.809 (0.007)&
    16.818 (0.009)& 17.418 (0.016)\\ 258.67& CTIO 1.5-m &
    Aldering &\nodata & 16.864 (0.005)& 16.801 (0.005)& 17.421
    (0.012)\\ 259.80& Lick 40-inch& Quimby &17.123 (0.041)&
    16.905 (0.029)& 16.814 (0.024)& \nodata \\ 265.57& YALO &
    Service &17.311 (0.009)& 17.108 (0.011)& 17.095 (0.012)& 17.783
    (0.026)\\ 266.68& CTIO 0.9-m & Strolger &17.412 (0.013)&
    17.210 (0.017)& 17.172 (0.019)& 17.875 (0.030)\\ 272.72&
    CTIO 0.9-m & Germany &17.905 (0.016)& 17.533 (0.054)& 17.521
    (0.020)& 18.191 (0.049)\\ 274.73& CTIO 0.9-m & Strolger
    &18.146 (0.031)& 17.624 (0.032)& 17.610 (0.017)& 18.234 (0.035)\\
    275.71& CTIO 0.9-m & Strolger &18.245 (0.020)& 17.625
    (0.048)& 17.615 (0.049)& 18.236 (0.045)\\ 275.82& Lick
    40-inch& Quimby &18.197 (0.055)& \nodata & 17.628 (0.026)& \nodata
    \\ 276.72& CTIO 1.5-m & Smith + Strolger&18.298 (0.009)&
    17.697 (0.008)& 17.637 (0.012)& 18.182 (0.020)\\ 277.69&
    CTIO 1.5-m & Smith + Strolger&18.365 (0.009)& 17.725 (0.008)&
    17.640 (0.009)& 18.160 (0.025)\\ 280.87& Lick 40-inch&
    Gates &18.680 (0.171)& 17.798 (0.066)& 17.613 (0.062)& \nodata \\
    284.48& Lick 40-inch& Quimby &18.907 (0.051)& 17.948
    (0.036)& \nodata & \nodata \\ 284.62& YALO & Service
    &18.917 (0.019)& 17.957 (0.015)& 17.683 (0.018)& 17.947 (0.044)\\
    285.46& JKT 1-m &M\'{e}ndez + Blanc&\nodata &
    17.991 (0.006)& \nodata & 17.893 (0.014)\\ 287.63& YALO &
    Service &19.110 (0.008)& 18.096 (0.009)& 17.703 (0.005)& 17.923
    (0.015)\\ 288.82& Lick 40-inch& Gates &19.219 (0.138)&
    \nodata & 17.746 (0.120)& \nodata \\ 291.68& CTIO 1.5-m &
    Smith + Strolger&19.378 (0.020)& 18.277 (0.035)& 17.855 (0.019)&
    17.847 (0.021)\\ 292.61& CTIO 1.5-m & Smith +
    Strolger&19.396 (0.044)& 18.348 (0.022)& 17.889 (0.024)& 17.854
    (0.060)\\ 297.46& CTIO 1.5-m & Strolger &19.629 (0.080)&
    18.637 (0.080)& 18.209 (0.022)& 18.161 (0.106)\\ 303.67&
    CTIO 1.5-m & Strolger &19.766 (0.031)& 18.936 (0.100)& \nodata &
    \nodata \\ 304.65& CTIO 1.5-m & Strolger &19.743 (0.016)&
    18.948 (0.011)& 18.494 (0.007)& 18.543 (0.036)\\ 313.60&
    YALO & Service &19.856 (0.005)& 19.236 (0.007)& 18.891 (0.015)&
    19.093 (0.081)\\ \enddata
    \tablecomments{JD is 2,451,000 $+$ \ldots}
\end{deluxetable}

\begin{deluxetable}{ccccc}
    \tablewidth{0pc}
     \tablenum{3}

    \tablecaption{Peak and Decline Rate Data\label{table:peaks}}
    \tablehead{
    \colhead{Filter} & \colhead{JD$_{max}$} & \colhead{$m_{max}$} 
    & \colhead{$M_{max}$} & \colhead{$\Delta$$m_{15}$}}
    \startdata
    
    B&254.7(0.3)&16.88(0.01)&-19.45(0.11)&0.81(0.03)\\
    V&255.6(0.3)&16.81(0.01)&-19.50(0.11)&0.66(0.04)\\
    R&256.0(0.3)&16.79(0.01)&-19.38(0.11)&0.62(0.04)\\
    I\tablenotemark{a}&251.8(4.2)&17.24(0.08)&-18.97(0.12)&0.61(0.09)\\
    \enddata \tablecomments{JD$_{max}$ is the date of maximum, and
    $m_{max}$ is the apparent magnitude at maximum.  $M_{max}$ is the
    absolute magnitude at maximum (assuming $H_{o}=65$ km/sec/Mpc),
    K-corrected, and corrected for time dilation and Galactic
    extinction (Schlegel, Finkbeiner, \& Davis 1998). 
    $\Delta$$m_{15}$ is decline rate of passband lightcurve from
    passband maximum, also K-corrected and corrected for time
    dilation.  JD is 2,451,000 $+$ \ldots} \tablenotetext{a}{{\it
    I}-band observations clearly do not pass though the peak, thus
    making it difficult to determine the maximum with great
    certainty.}
\end{deluxetable}

\begin{deluxetable}{ccccccc}
    \tablenum{4}
    \tablewidth{0pc} \tablecaption{IR Local Standard
    Indices\label{table:localir}} \tablehead{ \colhead{Star} &
    \colhead{J$_{s}$} & \colhead{H} & \colhead{K$_{s}$} &
    \colhead{N$_{Js}$} & \colhead{N$_{H}$} & \colhead{N$_{Ks}$}}
    \startdata
    
    2&15.520(0.015)&15.189(0.018)&15.182(0.035)&5&10&11\\
    3&17.236(0.036)&16.854(0.038)&16.813(0.070)&5&10&1\\
    4&16.095(0.017)&15.646(0.021)&15.678(0.048)&4&10&11\\
    18&17.113(0.032)&16.455(0.031)&16.357(0.057)&4&10&10\\
    19&15.671(0.016)&15.125(0.019)&14.863(0.032)&5&10&11\\ \enddata
    \tablecomments{Local photometric sequence in IR with mean error
    and number of observations.  Stars are numbered using the same
    system used in the optical observations.}
\end{deluxetable}    
    
 \begin{deluxetable}{cccccc}
    \tablewidth{0pc}
    \tablecaption{IR PSF Photometry\label{table:irobs}}
    \tablenum{5}
    \tablehead{
    \colhead{JD} & \colhead{Telescope} & \colhead{Observer} &
    \colhead{J$_{s}$} & \colhead{H} & \colhead{K$_{s}$}}
    \tabletypesize{\footnotesize}
    \startdata 
    
254.66&LCO 1-m/C40IRC&Phillips, Roth&17.719(0.035)&18.031(0.039)&17.525(0.075)\\
255.68&LCO 1-m/C40IRC&Roth&17.853(0.049)&\nodata&17.481(0.065)\\
256.67&LCO 1-m/C40IRC&Roth&17.824(0.033)&18.000(0.048)&17.566(0.197)\\
257.66&LCO 1-m/C40IRC&Roth&17.916(0.041)&18.305(0.053)&17.739(0.097)\\
258.71&LCO 1-m/C40IRC&Roth&17.914(0.035)&18.129(0.051)&17.735(0.092)\\
259.67&LCO 1-m/C40IRC&Roth&18.058(0.043)&18.193(0.045)&17.413(0.093)\\
260.67&LCO 1-m/C40IRC&Roth&18.235(0.043)&18.319(0.058)&17.645(0.078)\\
262.68&LCO 2.5-m/C100IRC&Galaz&18.361(0.045)&18.167(0.057)&17.589(0.060)\\
264.57&LCO 2.5-m/C100IRC&Galaz&18.495(0.041)&18.123(0.061)&17.558(0.050)\\
270.62&LCO 1-m/C40IRC&Muena&19.336(0.129)&18.263(0.091)&17.937(0.114)\\
272.61&LCO 2.5-m/CIRSI&Marzke, Persson&19.118(0.038)&\nodata&\nodata\\
273.53&LCO 2.5-m/CIRSI&Marzke, Persson&\nodata&18.406(0.033)&\nodata\\
275.49&LCO 2.5-m/CIRSI&Marzke, Persson, Phillips&18.965(0.072)&\nodata&\nodata\\
288.59&LCO 1-m/C40IRC&Hamuy&18.658(0.055)&18.051(0.063)&\nodata\\
290.56&LCO 1-m/C40IRC&Phillips&18.527(0.045)&18.177(0.082)&18.021(0.100)\\
294.59&LCO 1-m/C40IRC&Roth&18.551(0.056)&18.031(0.056)&\nodata\\
301.47&LCO 2.5-m/CIRSI&McCarthy&\nodata&18.528(0.039)&\nodata\\
302.5&LCO 2.5-m/CIRSI&McCarthy, Phillips&19.422(0.047)&\nodata&\nodata\\
304.48&VLT/ISAAC&Hamuy&19.684(0.129)&18.722(0.123)&18.853(0.128)\\

    \enddata \tablecomments{{\em J$_{s}$ H K$_{s}$} PSF photometry
    of SN~1999aw with photometric error. JD is 2,451,000 $+$ \ldots}
\end{deluxetable}

\begin{deluxetable}{cccc}
    \tablewidth{0pc}
    \tablenum{6}
    \tablecaption{Peak Luminosity and $^{56}$Ni Masses of Type Ia 
    Supernovae\label{table:masses}}
    \tablehead{
    \colhead{SN} & \colhead{$\Delta$$m_{15}(B)$} & \colhead{Log 
    L$_{bol}$ (erg s$^{-1}$)} & \colhead{M$_{Ni}$ (M$_{\odot}$)}}
    \startdata

    1989B&1.20&43.06&0.57\\ 1991T\tablenotemark{b}&0.97&43.23&0.84\\
    1991bg&1.85&42.32&0.11\\ 1992A&1.33&42.88&0.39\\
    1992bc&0.87&43.22&0.84\\ 1992bo&1.73&42.91&0.41\\
    1994D&1.46&42.91&0.41\\ 1994ae&0.95&43.04&0.55\\
    1995D&0.98&43.19&0.77\\ {\bf 1999aw}&{\bf 0.81}&{\bf 43.18}&{\bf 0.76}\\
    \enddata

    \tablecomments{Bolometric luminosity at maximum(in erg s$^{-1}$;
    estimated from UBVRI lightcurves) and estimated $^{56}$Ni mass
    (in M$_{\odot}$) of SN~1999aw compared to values of Contardo et
    al.  2000.  The $\Delta$$m_{15}(B)$ is also listed.  }
    \tablenotetext{b}{Contardo et al.  assume a distance modulus to
    SN~1991T of 31.07.  However, Saha et al.  (2001) have found a
    distance modulus of 30.74.  The luminosity and nickel mass are
    based on this new value.  }
\end{deluxetable}

\begin{deluxetable}{c}
     \tablewidth{0pc}
    \tablenum{A7}
     \tablecaption{Photometric Transformation Equations\label{table:eqn}}
     \tablehead{}
     \startdata
     $B-b=A1+A2\times (b-v)+A3\times X$\\
     $V-v=B1+B2\times (b-v)+B3\times X$\\  
     $R-r=C1+C2\times (v-i)+C3\times X$\\   
     $I-i=D1+D2\times (v-i)+D3\times X$\\
     \enddata
 \end{deluxetable}

\begin{deluxetable}{ccccc}
    \tablenum{A8}
    \tablewidth{0pc} \tablecaption{Coefficients to the Transformation
    Equations in Table~\ref{table:eqn}\label{table:terms}} \tablehead{
    \colhead{Coefficient} & \colhead{Telescope} & \colhead{1} &
    \colhead{2} & \colhead{3}} \startdata A&CTIO 0.9-m&3.477
    (0.027)&0.092 (0.005)&0.246 (0.029)\\ &CTIO 1.5-m&2.416
    (0.010)&0.094 (0.003)&0.205 (0.025)\\ B&CTIO 0.9-m&3.142
    (0.026)&$-$0.015 (0.002)&0.125 (0.017)\\ &CTIO 1.5-m&2.086
    (0.020)&$-$0.015 (0.002)&0.119 (0.027)\\ C&CTIO 0.9-m&3.154
    (0.026)&$-$0.006 (0.004)&0.085 (0.022)\\ &CTIO 1.5-m&2.062
    (0.031)&$-$0.005 (0.004)&0.071 (0.018)\\ D&CTIO 0.9-m&3.923
    (0.024)&$-$0.016 (0.004)&0.048 (0.019)\\ &CTIO 1.5-m&2.901
    (0.011)&$-$0.015 (0.004)&0.042 (0.027)\\
    \tablecomments{Coefficients should be read by row first, and then by
	column.  Example: the coefficient in row A, column number 1 is
	coefficient A1 for equation in Table~\ref{table:eqn}.  They
	are given for both the CTIO 0.9-meter and 1.5-meter telescopes
	along with the mean error.}
	\enddata
    \end{deluxetable}


\begin{references}
     
     \reference{} Ardeburg, A., \& de Groot, M. 1973, \aap, 28, 295
     
     \reference{} Arnett, W. D. 1982, \apj, 253, 785
     
     \reference{} Aldering, G. 2000, ``Type Ia Supernovae and Cosmic 
     Acceleration,'' AIP Conference Proceeding: Cosmic Explorations, ed. 
     S. S. Holt \& W. W. Zhang, Woodbury, New York: American Institute 
     of Physics. 
     
     \reference{} Bessell, M. S. 1979, \pasp, 91, 589
     
     \reference{} Bessell, M. S., \& Brett, J. M. 1988, \pasp, 100, 1134 
     
     \reference{} Bessell, M. S. 1990, \pasp, 102, 1181
     
     \reference{} Branch, D., Fisher, A., \& Nugent, P. 1993, \aj, 
     106, 2383
     
     \reference{} Branch, D., \& Patchett, B. 1973, \mnras, 161, 71
     
     \reference{} Cadonau, R., \& Leibundgut, B. 1990, \aaps, 82, 145
     
     \reference{} Cardelli, J. A., Clayton, G. C., \& Mathis, J. S.
     1989, \apj, 345, 245
     
     \reference{} Cohen, J. G., Persson, S. E., Elias, J. H., \& Frogel, 
     J. A. 1981, \apj, 249, 481
     
     \reference{} Contardo, G., Leibundgut, B., and Vacca, W. D. 
     2000, \aap, 359, 876
     
     \reference{} Elias, J. H., Matthews, K., Neugebauer, G., \& 
     Persson, E. 1985b, \apj, 296, 379
     
     \reference{} Filippenko, A., et al. 1992, \aj, 104, 4, 1556


     
     \reference{} Garnavich, P., et al. 2001, preprint (astro-ph/0105490)
     
     \reference{} Hamuy, M., Phillips, M. M., Maza, J., Wischnjewsky,
     M., Uomoto, A., Landolt, A. U., \& Khatwani, R. 1991, \aj, 102,
     208
     
     \reference{} Hamuy, M., Phillips, M. M., Wells, Lisa A., \& Maza, 
     Jose 1993, \pasp, 105, 689, 787
     
     \reference{} Hamuy, M., Phillips, M. M., Maza, J., Suntzeff, 
     N. B., Schommer, R. A., Aviles, R. 1995, \aj, 109, 1669, 1
     
     \reference{} Hamuy, M., Phillips, M. M., Suntzeff, N. B.,
     Schommer, R. A., Maza, J., \& Aviles, R. 1996a, \aj, 112, 2391    
     
     \reference{} Hamuy, M., Phillips, M. M., Suntzeff, N. B.,
     Schommer, R. A., Maza, J., \& Aviles, R. 1996b, \aj, 112, 2398
     
     \reference{} Hamuy, M., et al. 1996c, \aj, 112, 2408
     
     \reference{} Hamuy, M., Phillips, M. M., Suntzeff, N. B.,
     Schommer, R. A., Maza, J., Smith, R. C., Lira, P., \& Aviles, R.
     1996d, \aj, 112, 2438
     
     \reference{} Hamuy, M., Trager, S. C., Pinto, P. A., Phillips, M.
     M., Schommer, R. A., Ivanov, V., \& Suntzeff, N. B. 2000, \aj,
     120, 1479
     
     \reference{} Henry, R. B. C., \& Worthey, G. 1999, \pasp, 111, 919

     \reference{} Hoeflich, P., Wheeler, J.~C., Thielemann, F.~K. 
     1998, \apj, 495, 617

     \reference{} Ivanov, V. D., Hamuy, M., \& Pinto, P. A. 2000,
     \apj, 542, 588
     
     \reference{} Krisciunas, K., Hastings, N. C., Loomis, K.,
     McMillan, R., Rest, A., Riess, A. G., \& Stubbs, C. 2000, \apj,
     539, 658
 
     \reference{} Landolt, A. U. 1992, \aj, 104, 340
     
     \reference{} Lee, T. A., Wamsteker, W., Wisniewski, W. Z., \&
     Wdowiak, T. J. 1972, \apj, 177, L59
     
     \reference{} Leibundgut, B., et al. 1993, \aj, 105, 301
     
     \reference{} Li, W. D., Filippenko, A. V., Treffers, R. R.,
     Riess, A. G., Hu, J., \& Qiu, Y. 2001, \apj, 546, 734
     
     \reference{} Lira, P. 1995, in Masters thesis, Universidad de 
     Chile. 

     \reference{} Makarova, L. 1999, \aaps, 139, 491
     
     \reference{} Mazzali, P. A., Nomoto, K., Cappellaro, E.,
     Nakamura, T., Umeda, H., \& Iwamoto, K. 2001, \apj, 547, 988

     \reference{} Minkowski, R. 1939, \apj, 89, 156
     
     \reference{} Minkowski, R. 1940, \pasp, 52, 206
     
     \reference{} Nugent, P., Phillips, M. M., Baron, E., Branch, D.,
     \& Hauschildt, P. 1995a, \apj, 455, L147
     
     \reference{} Nugent, P., Branch, D., Baron, E., Fisher, A.,
     Vaughan, T., \& Hauschildt, P. H. 1995b, \prl, 75, 394
     
     \reference{} Nugent, P., Aldering, G. 2000, ``The Spring 1999
     Nearby Supernovae Campaign'', The Greatest Explosions Since the
     Big Bang: Supernovae and Gamma-ray Bursts, Space Telescope
     Science Institute Symposium, May 1999, ed.  M. Livio, N. Panagia,
     \& K. Sahu, Baltimore: Space Telescope Science Institute.

     \reference{} Nugent, P. et al. 2002, in preparation

     \reference{} Perlmutter, S., et al. 1997, \apj, 483, 565
 
     \reference{} Perlmutter, S., et al. 1998, \nat, 391, 51
     
     \reference{} Perlmutter, S., \& Supernova Cosmology Project 
     1999, \apj, 517, 565
     
     \reference{} Persson, S. E., Murphy, D. C., Krzeminski, W., 
     Roth, M., \& Rieke, M. J. 1998, \aj, 116, 2475
     
     \reference{} Phillips, M. M., Wells, L. A., Suntzeff, N. B.,
     Hamuy, M., Leibundgut, B., Kirshner, R. P., \& Foltz, C. B. 1992,
     \aj, 103, 1632
     
     \reference{} Phillips, M. M. 1993, \apj, 413, L105
     
     \reference{} Phillips, M. M., Lira, P., Suntzeff, N. B.,
     Schommer, R. A., Hamuy, M., \& Maza, J. 1999, \aj, 118, 1766
     
     \reference{} Phillips, M. M., et al., in preparation
     
     \reference{} Pinto, P. A., \& Eastman, R. G. 2000a, \apj, 
     530, 744
     
     \reference{} Pinto, P. A., \& Eastman, R. G. 2000b, \apj, 
     530, 757
     
     \reference{} Pskovskii, Y. P. 1969, Soviet Astronomy, 12, 750
     
     \reference{} Riess, A. G., Press, W. H., \& Kirshner, R. P., 
     1996, \apj, 473, 88
     
     \reference{} Riess, A. G., et al. 1999, \aj, 118, 2675
     
     \reference{} Saha, A., Labhardt, L., Schwengeler, H., Macchetto,
     F. D., Panagia, N., Sandage, A., \& Tammann, G. A. 1994, \apj,
     425, 14
     
     \reference{} Saha, A., Sandage, A., Thim, F., Labhardt, L.,
     Tammann, G. A., Christensen, J., Panagia, N., \& Macchetto, F. D.
     2001, \apj, 551, 973
     
     \reference{} Schlegel, D., Finkbeiner, D., \& Davis, M. 1998, 
     \apj, 500,525
     
     \reference{} Stetson, Peter B. 1992, Astronomical Data Analysis 
     Software and Systems, 25, 297
     
     \reference{} Strolger, L. -G., Smith, R. C., Clocchiatti, A., 
     Phillips, M. M., Suntzeff, N. B., \& NGSS Project Team 2000, 
     American Astronomical Society Meeting 197, \#81.01
     
     \reference{} Strolger, L. -G., et al. 2003a, in preparation
     
     \reference{} Strolger, L. -G., et al. 2003b, in preparation
     
     \reference{} Suntzeff, N. B., Bouchet, P. 1990, \aj, 99, 650
     
     \reference{} Suntzeff, N. B. 1996, in: McCray R., \& Wang, Z. 
     (eds.), AU Colloquium 145: Supernovae and Supernovae Remnants, 
     Cambridge: University Press, p.41
     
     \reference{} Wells, L., et al. 1994, \aj, 108, 2233

 \end{references}
\end{document}